\documentclass[aps,twocolumn,superscriptaddress]{revtex4-2}
\usepackage{graphicx}
\usepackage{dcolumn}
\usepackage{bm}
\usepackage{xcolor}

\begin{document}
	
	\preprint{APS/123-QED}

\title{Achieving 5 \% $^{13}$C nuclear spin hyperpolarization in high-purity diamond at room temperature and low field}

\author{Vladimir V. \surname{Kavtanyuk}}
\affiliation{Quantum magnetic sensing group, Korea Research Institute of Standards and Science, Daejeon 34113, Republic of Korea}
\author{Changjae \surname{Lee}}
\affiliation{School of Biomedical Engineering, Korea University, Seoul 02842, Republic of Korea}
\affiliation{Department of Chemistry, Korea Military Academy, Seoul 01805, Republic of Korea}
\author{Keunhong \surname{Jeong}}
\email{doas1mind@kma.ac.kr}
\affiliation{Department of Chemistry, Korea Military Academy, Seoul 01805, Republic of Korea}
\author{Jeong Hyun  \surname{Shim}}
\email{jhshim@kriss.re.kr}
\affiliation{Quantum magnetic sensing group, Korea Research Institute of Standards and Science, Daejeon 34113, Republic of Korea}
\affiliation{Department of Applied Measurement, University of Science and Technology, Daejeon 34113, Republic of Korea}

\date{\today}%

	\begin{abstract}
Optically polarizable nitrogen-vacancy (NV) center in diamond enables the hyperpolarization of $^{13}$C nuclear spins at low magnetic field and room temperature. However, achieving a high level of polarization comparable to conventional dynamic nuclear polarization has remained challenging.  Here we demonstrate that, at below 10 mT, a $^{13}$C polarization of 5\% can be obtained, equivalent to an enhancement ratio over $7 \times 10^6$.  We used high-purity diamond with a low initial nitrogen concentration ($<$ 1 ppm), which also results in a long storage time exceeding 100 minutes. By aligning the magnetic field along [100], the number of NV spins participating in polarization transfer increases fourfold. We conducted a comprehensive optimization of field intensity and microwave (MW)  frequency-sweep parameters for this field orientation. The optimum MW sweep width suggests that polarization transfer occurs primarily to bulk $^{13}$C spins through the integrated solid effect followed by nuclear spin diffusion. 
	\end{abstract}

	\maketitle

\section{Introduction}

Dynamic nuclear polarization (DNP) effectively enhances nuclear spin polarization to a decent level \cite{Goldman1970,Abragam1976,Ardenkjar2003,Ardenkjaer2015,Yoon2018,Corzilius2020,Eills2023}. It capitalizes on the transfer of spin polarization from electrons to nuclei, resulting in hyperpolarization that surpass thermal polarization by several orders of magnitude. But, DNP necessarily relies on cryogenic temperatures and high magnetic fields. 

In contrast, optical hyperpolarization based on negatively charged nitrogen vacancy (NV) centers in diamond can be performed at room temperature and low magnetic fields \cite{Fischer2013,Wang2013,Alvarez2015,King2015,Scheuer2016}. The optically induced polarization of NV electron spins  is transferred to $^{13}$C nuclear spins under microwave (MW) irradiation.  Various NV-based optical hyperpolarization methods have been explored, including continuous-wave (CW) and pulsed MW irradiations \cite{Fischer2013,Wang2013,Alvarez2015,King2015,Scheuer2016,Ajoy2018, Parker2019, Scheuer2020}. The CW method has been extended to repetitive irradiation of frequency-swept MW. The frequency sweep induces nuclear spin-selective adiabaticity during Landau-Zener (LZ) transition within a pair of NV and local $^{13}$C spins. Under laser illumination, the nuclear spin selectivity results in hyperpolarization of the local nuclear spins. This spin ratchet process is of particular interest as it applies to diamond powders, where NV spins are randomly oriented \cite{Ajoy2018,Ajoy2020}.

 \begin{figure}[b!]
	\includegraphics[width=7.8 cm]{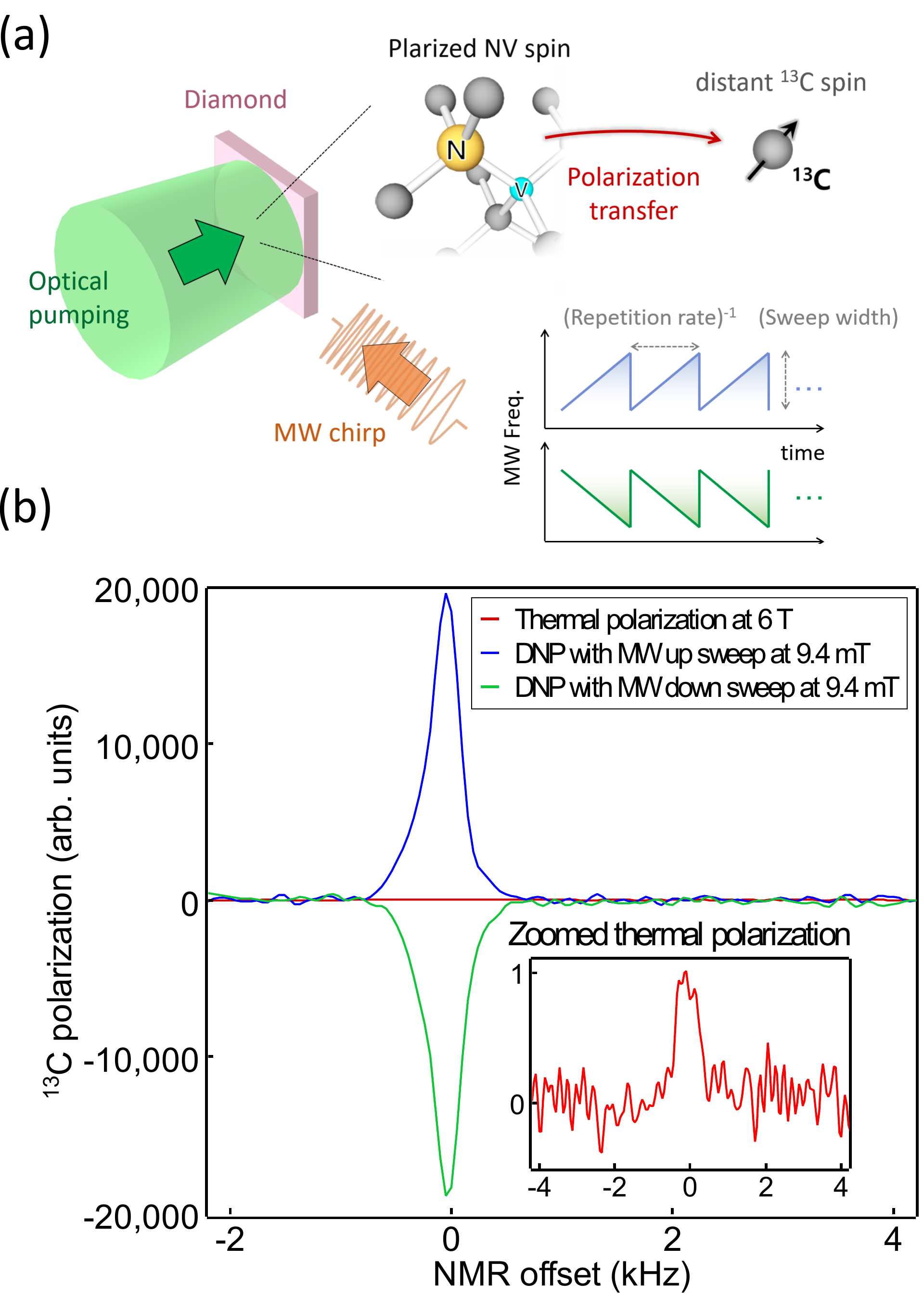}
	\caption{(a) Schematic of the optical hyperpolarization of $^{13}$C spins in diamond; (b) Hyperpolarization and thermal polarization of $^{13}$C. The blue (MW sweeps are applied from up to down) and green (MW sweeps are applied from down to up) lines represent real part of Fourier transform signals from 5 \% of $^{13}$C polarization obtained at 9.4 mT. The red line represents real part of FT from the thermal polarization at 6 T.
	\label{fig1}}
\end{figure}

\begin{figure*}[t!]
	\includegraphics[width=\textwidth]{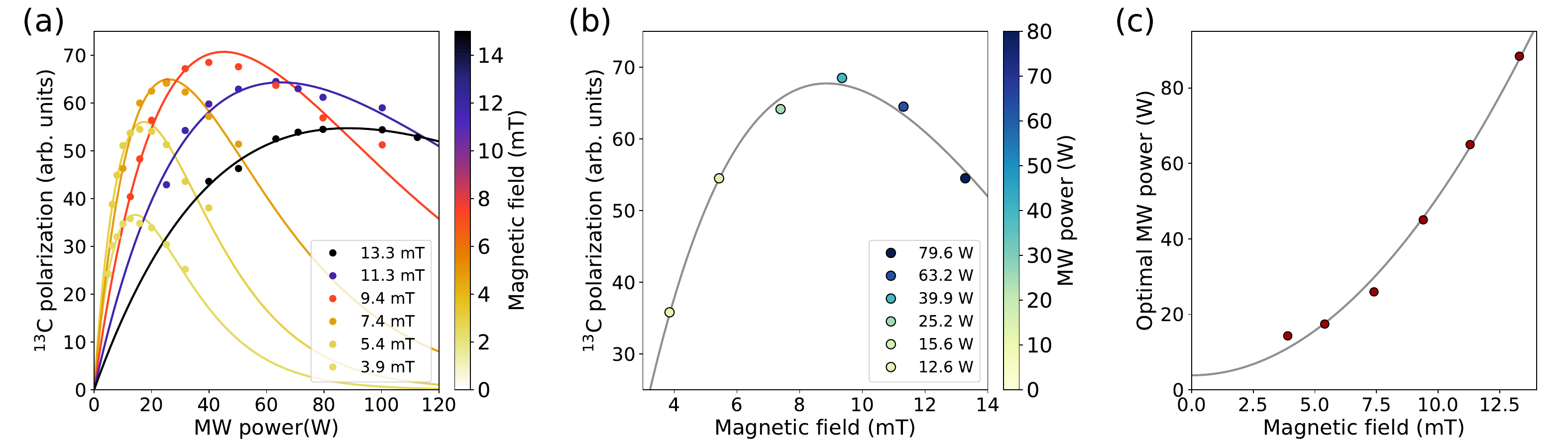}
	\caption{(a) $^{13}$C polarization vs MW power at different magnetic fields; (b) $^{13}$C polarization vs magnetic field with MW power set to the maximum polarization values; (c) Optimal MW power as a function of magnetic field, where the data points show a quadratic relationship between the optimal MW power and applied magnetic field. All these measurements are done with a sweep width of 100 MHz and a sweep rate of 10 MHz/ms.
		\label{fig3}}
\end{figure*}

Several studies have aimed to further enhance $^{13}$C polarization. One approach involved overcoming instrumental limitation: using multiple lasers with a combined power exceeding 20 W significantly increased the polarization injection rate \cite{Sarkar2022}. Another approach improved the quality of diamond via thermal annealing, which helped mitigate the influences of paramagnetic impurities \cite{Gierth2020}. Despite these efforts, NV-based optical hyperpolarization still lacks a comparative advantage over conventional DNP in terms of the achievable level of polarization.

In this study, we present a different approach by using a high-purity diamond with a considerably low nitrogen (N) concentration. Substitutional N atoms (N$_{\mathrm{S}}^0$) are paramagnetic and the most abundant source of decoherence, shortening T$_2$ times of NV and $^{13}$C spins \cite{Bauch2020}. Reducing the N concentration potentially improves the efficiency of polarization transfer by mitigating the effects of fluctuating local magnetic fields, which can interfere with the external field driving the LZ transition \cite{Nalbach2010}. However, the limited conversion yield from N to NV results in a reduced number of polarization sources. To address this issue, we applied the external field along the [100] axis, effectively increasing the number of NV spins contributing to the optical hyperpolarization fourfold. We conducted a comprehensive optimization of key parameters, including magnetic field strength, microwave power, and sweep characteristics, which are crucial for maximizing the efficiency of polarization transfer. As a result, we were able to obtain 5 \% polarization of $^{13}$C nuclei (Fig.~1(b)). Notably, this high polarization level was accompanied by an exceptionally long storage time exceeding 100 minutes.

 \section{Experimental methods}
We utilized a CVD-grown single crystal diamond weighing 15 mg with [100] surface orientation (from Element Six). This orientation was aligned with the direction of the magnetic field during our measurements. The diamond possesses a natural abundance of $^{13}$C nuclei (1.1\%), along with approximately 0.2 ppm of N$_{\mathrm{S}}^0$ and 0.3 ppm of NV centers. To estimate N$_{\mathrm{S}}^0$ concentration, we employed electron spin resonance techniques (see Fig.~6 of the supplementary material). The polarization of $^{13}$C in the diamond was evaluated by referencing the thermal polarization of a diamond crystal (42 mg), as detailed in our previous work \cite{molecules27051700}. Our instrumentation polarizes $^{13}$C nuclei in diamonds in the region of low magnetic fields (ranging from 1 to 50 mT), with subsequent NMR readouts performed at the center of  a 6 T superconducting magnet. To facilitate rapid transitions between the two distinctive magnetic regions, a swift shuttling device is installed, enabling the transfer of the diamond within 1.5 seconds (Fig.~4 of the supplementary material). A detail description of our experimental setup is provided in our previous publication \cite{molecules27051700}.

\section{Results and discussion}
\subsection{System and parameters}
For a NV concentration ($n$) of 0.3 ppm, the nearest-neighbor distance ($r_{\mathrm{NN}}$) is 16.7 nm (= $0.62 n^{-1/3}$). The $^{13}$C spin linewidth appears to be 1 kHz (Fig.~1(b)), which corresponds to the width of the spectral packet associated with nuclear spin diffusion. Given that the hyperfine interaction for $^{13}$C spins separated by 2 nm is in the order of 10 kHz \cite{Abobeih2019}, the radius ($r_{c}$) at which the hyperfine interaction becomes comparable to the spectral width (1 kHz) is estimated to be approximately 5 nm(This transition is gradual, and the effects of both hyperfine interaction and $^{13}$C$-$$^{13}$C dipolar interaction coexist over a range of distances). It is important to note that even if the local $^{13}$C spins within $r_{c}$ are fully polarized, achieving 5 \% polarization is still not feasible. The indirect polarization of bulk nuclear spins through the diffusion, initiated by direct polarization transfer via the solid effect, is necessarily taken into account. We previously observed the hallmark of the solid effect overlaid on $^{14}$N hyperfine splitting using CW-MW \cite{molecules27051700}. The bulk polarization beyond $r_{c}$ grows exponentially \cite{Wenckebach2016}. Moreover, the diffusion length ($L = \sqrt{D \cdot T_{\mathrm{pol}}}$, where $D$ = 0.67 nm$^2$/s \cite{Parker2019} and T$_{\mathrm{pol}}$ = 10.4 minutes in Fig.~4(a)) amounts to 24 nm, which is sufficient to cover the $r_{\mathrm{NN}}$. Therefore, we propose that, under frequency-swept MW, the polarization transfer via the integrated solid-effect (ISE) \cite{Can2018}  (Fig.~1(a)), followed by the nuclear spin diffusion, predominates and discuss its consistency with our results accordingly. 

We choose [100] for magnetic field orientation. The spectral positions of the ODMR peaks for the four NV axes become identical  (see Fig.~5(a) of the supplementary material). As a result, all NV centers can participate in the optical hyperpolarization. For this field orientation, we investigated the optimum conditions, at which the $^{13}$C polarization reaches the maximum. We considered five parameters: laser power ($P_{\mathrm{laser}}$), MW power ($P_{mw}$), sweep width ($\Delta$), sweep rate ($\dot{\Delta}$), and magnetic field ($B$). 
Two reciprocal relationships between $P_{mw}$ and $B$, and another between $\Delta$ and $\dot{\Delta}$, were observed as described sequentially below. 

\subsection{Magnetic field and MW power}
We aimed to find the optimum magnetic field. However, while $B$ increased along the [100] orientation, the dependence of MW power changed accordingly, as shown in Fig.~2(a). We found that each curve fits well to a phenomenological equation of $^{13}$C polarization ($\epsilon$) as follows:
\begin{equation}
\epsilon \propto P_{mw} \cdot \exp(-\beta P_{mw}).
\label{Opt_MW}
\end{equation}
The optimum MW powers at different magnetic fields were obtained from this equation. Fig.~2(c) shows that the collected values form a quadratic curve as a function of $B$. Figure~\ref{fig3}(b) illustrates the variation in the maximum $^{13}$C polarization. As the magnetic field increase, $^{13}$C polarization increases up to 9.4 mT, after which it begins to decrease. This decrease is attributed to the reduction in the NV spin polarization due to ground-state mixing when the magnetic field has an off-axis component \cite{Tetienne2012, Ajoy2020}. 

\begin{figure}[t!]
	\includegraphics[width=\columnwidth]{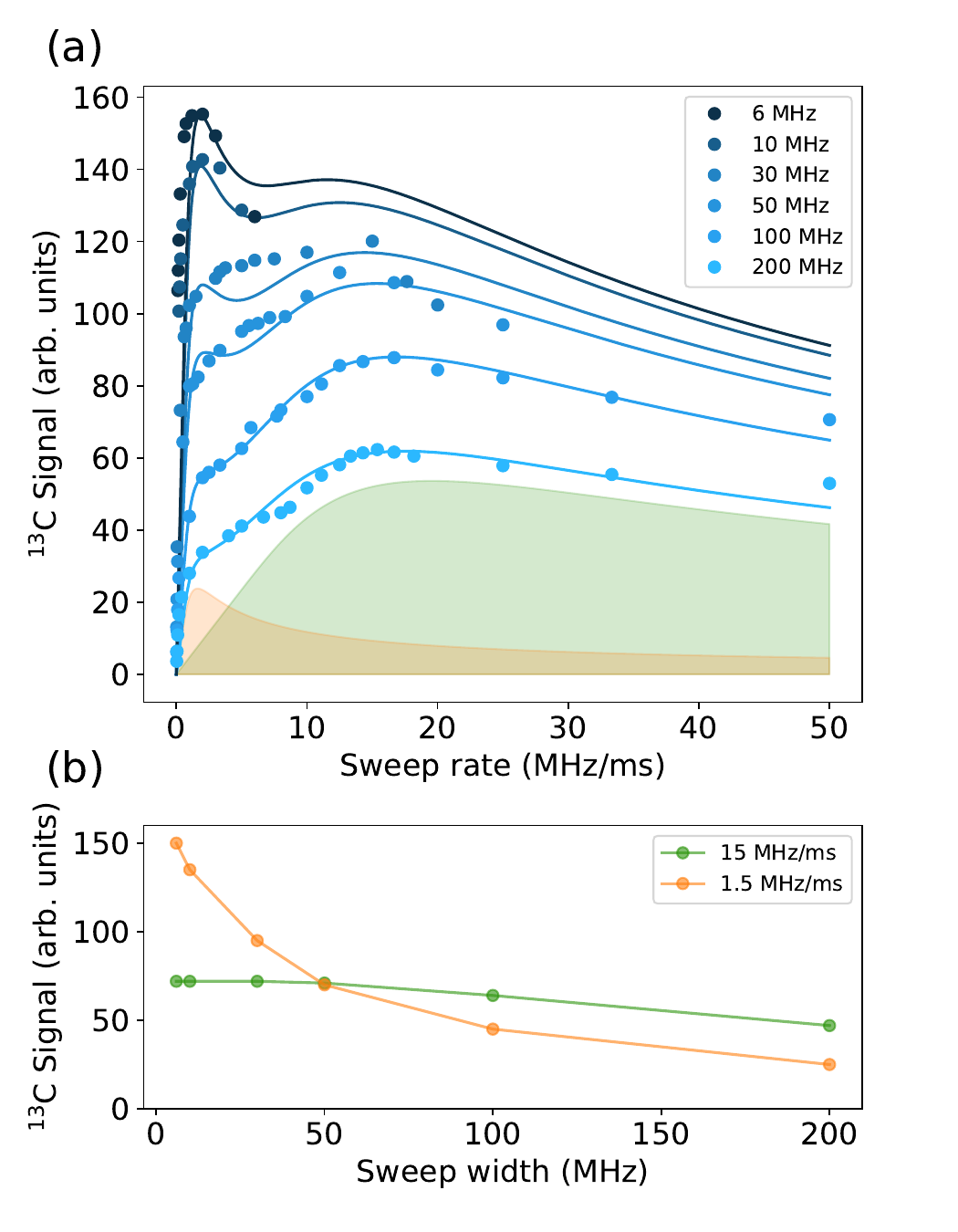}
	\caption{$^{13}$C polarization vs (a) sweeping rate at different MW ranges and (b) sweeping range at the optimal MW rates of 1.5 and 15 MHz/ms. These measurements are performed with MW power of 50 W and at a magnetic field of 9.4 mT.
		\label{fig4}}
\end{figure}

The correlation between $P_{mw}$ and $B$ shown in Fig.~2(a)~and~2(c) has not been reported previously. We suppose that such correlation is associated with the ISE process at low magnetic fields. In particular, Eq.~(\ref{Opt_MW}) can be interpreted as the product of the LZ transition probability and the polarization transfer efficiency. The LZ transition probability, $p$, is mathematically expressed as: 
\begin{equation}
  p = 1 - \exp\left(\frac{-c P_{mw}}{2 \pi \dot{\Delta}}\right).
  \label{LZ}
  \end{equation}
$p$ is proportional to $P_{mw}$ to the first-order approximation. $\dot{\Delta}$  is a fixed value in our experiment. The exponentially decaying term in Eq.~(\ref{Opt_MW}), originates from the ISE condition at low magnetic fields. During ISE, the polarization transfer occurs twice at detuned frequencies $\pm\delta f_{mw}$. $\delta f_{mw}$ is given by the solid effect condition \cite{Can2018}, 
\begin{equation}
\delta f_{mw} = \sqrt{\omega_{C}^2 - \Omega_{mw}^2} = \sqrt{(\gamma_C B)^2 - c P_{mw}},
\label{SE}
\end{equation} 
where $\omega_{C}$ is $^{13}$C resonance frequency. $ \Omega_{mw}$ corresponds to mw field intensity $B_{mw}$ as $\Omega_{mw} = \gamma_{e} B_{mw}$. $\Omega_{mw}^2$ has a linear relationship with $P_{mw}$, such that $\Omega_{mw}^2 = c P_{mw}$, which also appears in Eq.~(\ref{LZ}). The adiabatic inversion of electron spin leads to the integration of the transferred polarizations at $\pm\delta f_{mw}$. 
In high fields, typically $\omega_{C}$ is larger than $\Omega_{mw}$. However, at a low field near 10 mT, $\omega_{C}$ can be comparable to $\Omega_{mw}$. Thus, increasing $P_{mw}$ can significantly reduce $\delta f_{mw}$, and polarization transfer and adiabatic inversion may occur nearly simultaneously. Further increase in $P_{mw}$ invalidates the SE condition in Eq.~(\ref{SE}), making the ISE highly inefficient. This qualitative description elucidates two counteracting effects of increasing $P_{mw}$, which are embodied in Eq.~(\ref{Opt_MW}). 

The quadratic dependence of the optimal MW power shown in Fig.~2(c) can also be understood within the context of ISE.  Assuming that $\delta f_{mw}$ remains fixed while $B$ increases, one can infer from Eq.~(\ref{SE}) that $P_{mw}$ would have a margin for increase, and this margin depends quadratically on $B$.

\subsection{MW sweep width and rate}
Another reciprocal relationship was found between the two MW sweep parameters, $\Delta$ and $\dot{\Delta}$.  It has been reported that the optimal value exists for $\dot{\Delta}$, but its correlation with $\Delta$ has not been investigated. Figure~\ref{fig4}(a) reveals that, for a wide sweep width ($\Delta$ = 200 MHz), the max  $^{13}$C polarization was obtained at 15 MHz/ms ($\dot{\Delta}_{\mathrm{H}}$). This value is about 40 $\%$ of that in a previous work \cite{Ajoy2018}. We presume this difference can be attributed to the intrinsic properties of diamond and the orientation of magnetic field. Notably, the entire curve appears to contain an additional peak with a local maximum at 1.5 MHz/ms  ($\dot{\Delta}_{\mathrm{L}}$). 
As $\Delta$ reduced to 6 MHz, the peak at $\dot{\Delta}_{\mathrm{L}}$ increased rapidly. Due to the instrumental limitation in the repetition time, we could not obtain the full range curve for $\Delta$ = 6 MHz. Nonetheless, it is clear that the peak intensity at $\dot{\Delta}_{\mathrm{L}}$ overwhelms that at $\dot{\Delta}_{\mathrm{H}}$. We extracted the two peak intensities quantitatively by fitting the data with an equation:
\begin{equation}
\epsilon \propto \dot{\Delta}^c \left( 1 - \exp(- b / \dot{\Delta} ) \right).
\label{MWsweep}
\end{equation}
The left term stems from the LZ transition probability, which was also exploited in Eq.~(\ref{Opt_MW}). For the right term, $\dot{\Delta}^c$, we don't have a concrete reason for its adoption, but, we found that the exponent $c$ ranging from 1.5 to 1.7 fit the data  reasonably well. With the two peaks centered at  $\dot{\Delta}_{\mathrm{L}}$ (orange) and $\dot{\Delta}_{\mathrm{H}}$ (green), we fitted the curves and obtained the peak intensities as a function of $\Delta$. Figure~\ref{fig4}(b) reveals that when $\Delta$ becomes lower than 50 MHz, the optimal sweep rate shifts from $\dot{\Delta}_{\mathrm{H}}$ to $\dot{\Delta}_{\mathrm{L}}$. 

Moreover, we observed that if $\Delta$ falls below 6 MHz, the $^{13}$C polarization decreases (not shown in Fig.~3). This decrease is attributed to the insufficient coverage of the NV spin's density of states, which one can estimate from the linewidth of ODMR spectrum. Due to a slight misalignment of the magnetic field, the ODMR width amounts to 6 MHz (see Fig.~5(a) of the supplementary material). We conclude that a minimal sweep width covering the ODMR spectrum is advantageous for enhanced polarization transfer. This sweep width condition sounds inconsistent with the spin ratchet mechanism \cite{Sarkar2022}, in which local nuclear spins are mostly involved. Instead, it supports our notion that the polarization is transferred to the bulk (distant) $^{13}$C spins, as illustrated in Fig.~1(a), by bypassing the local $^{13}$C spins locating outside the spectral coverage of the MW sweep and thereby remaining unaffected. A similar dependence of MW sweep width was found for the ISE \cite{Can2018}.

\begin{figure}[t!]
	\includegraphics[width=\columnwidth]{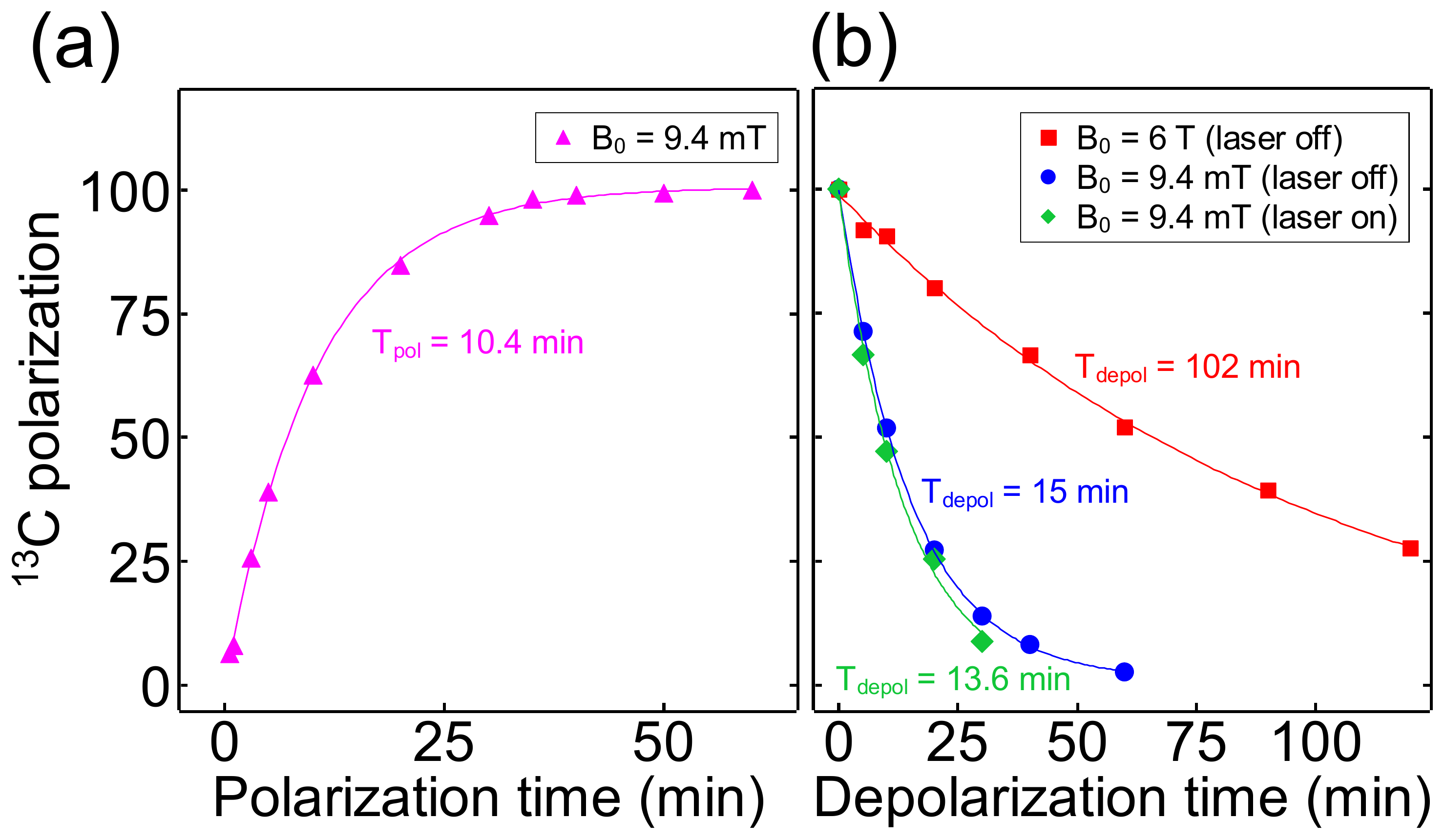}
	\caption{(a) $^{13}$C polarization build-up time measurement. The polarization saturation level is reached after 40 minutes; (b) Depolarization of $^{13}$C in the diamond at 6 T and at 9.4 mT. T$_{\mathrm{depol}}$ = 102 min at 6 T (laser off), T$_{\mathrm{depol}}$ = 15 min at 9.4 mT (laser off) and T$_{\mathrm{depol}}$ = 13.6 min at 9.4 mT (laser on).
		\label{fig2}}
\end{figure}

\subsection{Other factors}
The optimal magnetic field shown in Fig.~1(b) was obtained with a MW sweep rate of 10 MHz/ms, which is approximately in the middle between $\dot{\Delta}_{\mathrm{H}}$ and $\dot{\Delta}_{\mathrm{L}}$. This raises a question about the potential relationship between the optimal field and the MW sweep rate. We repeated the same experiments at different sweep rates, $\dot{\Delta}_{\mathrm{H}}$ and $\dot{\Delta}_{\mathrm{L}}$ (see Fig.~2 of the supplementary material). At $\dot{\Delta}_{\mathrm{L}}$, the optimal magnetic field has remained 9.4 mT. However, at $\dot{\Delta}_{\mathrm{H}}$, the max $^{13}$C polarization was achieved in a range of 9 to 13 mT. Therefore, the magnetic field of 9.4 mT in Fig. 2(b) may not represent the global optimum, but is highly effective in reaching the max polarization regardless of the MW sweep rate.

A previous work \cite{Sarkar2022} demonstrated that the optimal sweep rate or, more precisely repetition rate, increases significantly with the optical pumping power, though the MW power should be amplified accordingly. We conducted a similar test, measuring $P_{mw}$ dependence curves while increasing $P_{\mathrm{laser}}$. For each $P_{\mathrm{laser}}$, $P_{mw}$ was varied from 15 W to 65 W. The optimal $P_{mw}$ values are expected to show a notable increase. However, they remained nearly unaffected up to $P_{\mathrm{laser}}$ = 1 W. The origin of this discrepancy is unclear, but it is not related to the thermal heating of the diamond, because the temperature of the diamond was confirmed to be less than 320 K (see Fig.~3 and Fig.~7 of the supplementary material) \cite{Acosta2010}. The heating was highly suppressed by using a ceramic holder on which the diamond was firmly mounted with a heat-conducing epoxy. 

\subsection{$^{13}$C hyperpolarization and storage time}

Figure~\ref{fig1}(b) illustrates the maximum $^{13}$C polarization we achieved, reaching approximately 5\%. MW sweep for excitation was within the range from 2.740 to 2.746 GHz, both upward (blue line) and downward (green line). This sweep width covers one of the ODMR peaks located at a lower MW frequency. These measurements are conducted within a single run lasting 40 minutes, employing a magnetic field of 9.4 mT, MW power of 40 W, and laser power density of 30 mW/mm$^2$. The selection of these parameters follows the extensive optimization process described above sections. The thermal polarization of $^{13}$C (depicted in red in Fig.~\ref{fig1}) is acquired at 6 T from another diamond with 400 averaging scans. The enhancement factor of our $^{13}$C hyperpolarization over thermal polarization exceeds seven million (7 $\times$ 10$^6$). The ratio of the integrals of the hyperpolarized and thermally polarized $^{13}$C signals at 6 T is approximately 10,000.

Since we used a high-purity diamond, its exceedingly low impurity concentration leads to prolonged polarization build-up (T$_{\mathrm{pol}}$) and storage (T$_{\mathrm{depol}}$) times (Fig. 4(a),(b)). We measured T$_{\mathrm{depol}}$ at both low and high magnetic field conditions, as shown in Figure~\ref{fig2}(b). It was determined to be 100 minutes at 6 T (red) and 15 minutes at 9.4 mT (blue) in the absence of laser illumination. The laser illumination caused only a minor decrease in T$_{\mathrm{depol}}$, reducing it to 13.6 minutes (green). This indicates that the depolarization of $^{13}$C nuclear spins is nearly independent of the polarization cycles of NV spins. It has been known that the $^{13}$C spins proximal to NV spins experience strong depolarization under laser illumination \cite{Shim2013}.  The results in Fig.~4(b) further support that the majority of the optical hyperpolarization occurs in the bulk $^{13}$C spins interacting weakly with NV spins.  

\section{Conclusion}
In this study, we have demonstrated the achievement of 5 \% $^{13}$C hyperpolarization in diamond at room temperature and a magnetic field of 9.4 mT. The use of high-purity diamond with a low N concentration is believed to be the primary factor enabling such a high polarization. In addition, the parameters affecting the final $^{13}$C polarization were systematically optimized. During this process, two reciprocal relations emerged: one between magnetic field and MW power, and the other between MW sweep width and sweep rate. 
These correlations are partially consistent with the ISE at low magnetic fields. 
We suggest that the ISE becomes highly efficient in low N diamonds, where relaxations due to paramagnetic N impurities are significantly suppressed. A few of our findings still require further explanations, such as the emergence of an exceedingly small MW sweep rate. Nevertheless, we believe that this work provides a new perspective on NV-based optical hyperpolarization by highlighting the practical advantage of using low N concentration diamonds.

\begin{acknowledgments}
This work was supported by a grant (NP2023-0115) from Korea Research Institute of Standards and Science, Institute of Information and communications Technology Planning \& Evaluation (IITP) grants funded by the Korea government (MSIT) (2022-0-01026, RS-2023-00230717). C. Lee and K. Jeong were supported by Korea Research Institute for Defense Technology Planning and Advancement (KRIT) -Grant funded by Defense Acquisition Program Administration(DAPA) (KRIT-CT-23-031).
\end{acknowledgments}


\begin{thebibliography}{26}%
\makeatletter
\providecommand \@ifxundefined [1]{%
 \@ifx{#1\undefined}
}%
\providecommand \@ifnum [1]{%
 \ifnum #1\expandafter \@firstoftwo
 \else \expandafter \@secondoftwo
 \fi
}%
\providecommand \@ifx [1]{%
 \ifx #1\expandafter \@firstoftwo
 \else \expandafter \@secondoftwo
 \fi
}%
\providecommand \natexlab [1]{#1}%
\providecommand \enquote  [1]{``#1''}%
\providecommand \bibnamefont  [1]{#1}%
\providecommand \bibfnamefont [1]{#1}%
\providecommand \citenamefont [1]{#1}%
\providecommand \href@noop [0]{\@secondoftwo}%
\providecommand \href [0]{\begingroup \@sanitize@url \@href}%
\providecommand \@href[1]{\@@startlink{#1}\@@href}%
\providecommand \@@href[1]{\endgroup#1\@@endlink}%
\providecommand \@sanitize@url [0]{\catcode `\\12\catcode `\$12\catcode
  `\&12\catcode `\#12\catcode `\^12\catcode `\_12\catcode `\%12\relax}%
\providecommand \@@startlink[1]{}%
\providecommand \@@endlink[0]{}%
\providecommand \url  [0]{\begingroup\@sanitize@url \@url }%
\providecommand \@url [1]{\endgroup\@href {#1}{\urlprefix }}%
\providecommand \urlprefix  [0]{URL }%
\providecommand \Eprint [0]{\href }%
\providecommand \doibase [0]{https://doi.org/}%
\providecommand \selectlanguage [0]{\@gobble}%
\providecommand \bibinfo  [0]{\@secondoftwo}%
\providecommand \bibfield  [0]{\@secondoftwo}%
\providecommand \translation [1]{[#1]}%
\providecommand \BibitemOpen [0]{}%
\providecommand \bibitemStop [0]{}%
\providecommand \bibitemNoStop [0]{.\EOS\space}%
\providecommand \EOS [0]{\spacefactor3000\relax}%
\providecommand \BibitemShut  [1]{\csname bibitem#1\endcsname}%
\let\auto@bib@innerbib\@empty
\bibitem [{\citenamefont {Goldman}(1970)}]{Goldman1970}%
  \BibitemOpen
  \bibfield  {author} {\bibinfo {author} {\bibfnamefont {G.}~\bibnamefont
  {Goldman}},\ }\bibfield  {title} {\bibinfo {title} {Spin temperature and
  nuclear magnetic resonance in solids},\ }\href@noop {} {\bibfield  {journal}
  {\bibinfo  {journal} {Oxford Univ. Press}\ }\textbf {\bibinfo {volume} {1
  ed.}},\ \bibinfo {pages} {258} (\bibinfo {year} {1970})}\BibitemShut
  {NoStop}%
\bibitem [{\citenamefont {Abragam}\ and\ \citenamefont
  {Goldman}(1976)}]{Abragam1976}%
  \BibitemOpen
  \bibfield  {author} {\bibinfo {author} {\bibfnamefont {A.}~\bibnamefont
  {Abragam}}\ and\ \bibinfo {author} {\bibfnamefont {M.}~\bibnamefont
  {Goldman}},\ }\bibfield  {title} {\bibinfo {title} {Principles of dynamic
  nuclear polarization},\ }\href@noop {} {\bibfield  {journal} {\bibinfo
  {journal} {Rep. Prog. Phys.}\ }\textbf {\bibinfo {volume} {41}},\ \bibinfo
  {pages} {395} (\bibinfo {year} {1976})}\BibitemShut {NoStop}%
\bibitem [{\citenamefont {Ardenkjar-Larsen}\ \emph {et~al.}(2003)\citenamefont
  {Ardenkjar-Larsen}, \citenamefont {Fridlund}, \citenamefont {Gram},
  \citenamefont {Hansson}, \citenamefont {Hansson}, \citenamefont {Lerche},
  \citenamefont {Servin}, \citenamefont {Thaning},\ and\ \citenamefont
  {Golman}}]{Ardenkjar2003}%
  \BibitemOpen
  \bibfield  {author} {\bibinfo {author} {\bibfnamefont {J.~H.}\ \bibnamefont
  {Ardenkjar-Larsen}}, \bibinfo {author} {\bibfnamefont {B.}~\bibnamefont
  {Fridlund}}, \bibinfo {author} {\bibfnamefont {A.}~\bibnamefont {Gram}},
  \bibinfo {author} {\bibfnamefont {G.}~\bibnamefont {Hansson}}, \bibinfo
  {author} {\bibfnamefont {L.}~\bibnamefont {Hansson}}, \bibinfo {author}
  {\bibfnamefont {M.~H.}\ \bibnamefont {Lerche}}, \bibinfo {author}
  {\bibfnamefont {R.}~\bibnamefont {Servin}}, \bibinfo {author} {\bibfnamefont
  {M.}~\bibnamefont {Thaning}},\ and\ \bibinfo {author} {\bibfnamefont
  {K.}~\bibnamefont {Golman}},\ }\bibfield  {title} {\bibinfo {title} {Increase
  in signal-to-noise ratio of $>$10,000 times in liquid-state nmr},\
  }\href@noop {} {\bibfield  {journal} {\bibinfo  {journal} {Proc. Natl. Acad.
  Sci.}\ }\textbf {\bibinfo {volume} {100}},\ \bibinfo {pages} {10158–10163}
  (\bibinfo {year} {2003})}\BibitemShut {NoStop}%
\bibitem [{\citenamefont {Ardenkjaer-Larsen}\ \emph {et~al.}(2015)\citenamefont
  {Ardenkjaer-Larsen}, \citenamefont {Boebinger}, \citenamefont {Comment},
  \citenamefont {Duckett}, \citenamefont {Edison}, \citenamefont {Engelke},
  \citenamefont {Griesinger}, \citenamefont {Griffin}, \citenamefont {Hilty},
  \citenamefont {Maeda}, \citenamefont {Parigi}, \citenamefont {Prisner},
  \citenamefont {Ravera}, \citenamefont {Bentum}, \citenamefont {Vega},
  \citenamefont {Webb}, \citenamefont {Luchinat}, \citenamefont {Schwalbe},\
  and\ \citenamefont {Frydman}}]{Ardenkjaer2015}%
  \BibitemOpen
  \bibfield  {author} {\bibinfo {author} {\bibfnamefont {J.-H.}\ \bibnamefont
  {Ardenkjaer-Larsen}}, \bibinfo {author} {\bibfnamefont {G.~S.}\ \bibnamefont
  {Boebinger}}, \bibinfo {author} {\bibfnamefont {A.}~\bibnamefont {Comment}},
  \bibinfo {author} {\bibfnamefont {S.}~\bibnamefont {Duckett}}, \bibinfo
  {author} {\bibfnamefont {A.~S.}\ \bibnamefont {Edison}}, \bibinfo {author}
  {\bibfnamefont {F.}~\bibnamefont {Engelke}}, \bibinfo {author} {\bibfnamefont
  {C.}~\bibnamefont {Griesinger}}, \bibinfo {author} {\bibfnamefont {R.~G.}\
  \bibnamefont {Griffin}}, \bibinfo {author} {\bibfnamefont {C.}~\bibnamefont
  {Hilty}}, \bibinfo {author} {\bibfnamefont {H.}~\bibnamefont {Maeda}},
  \bibinfo {author} {\bibfnamefont {G.}~\bibnamefont {Parigi}}, \bibinfo
  {author} {\bibfnamefont {T.}~\bibnamefont {Prisner}}, \bibinfo {author}
  {\bibfnamefont {E.}~\bibnamefont {Ravera}}, \bibinfo {author} {\bibfnamefont
  {J.~V.}\ \bibnamefont {Bentum}}, \bibinfo {author} {\bibfnamefont
  {S.}~\bibnamefont {Vega}}, \bibinfo {author} {\bibfnamefont {A.}~\bibnamefont
  {Webb}}, \bibinfo {author} {\bibfnamefont {C.}~\bibnamefont {Luchinat}},
  \bibinfo {author} {\bibfnamefont {H.}~\bibnamefont {Schwalbe}},\ and\
  \bibinfo {author} {\bibfnamefont {L.}~\bibnamefont {Frydman}},\ }\bibfield
  {title} {\bibinfo {title} {Facing and overcoming sensitivity challenges in
  biomolecular nmr spectroscopy},\ }\href@noop {} {\bibfield  {journal}
  {\bibinfo  {journal} {Angew. Chem. Int. Ed.}\ }\textbf {\bibinfo {volume}
  {54}},\ \bibinfo {pages} {9162} (\bibinfo {year} {2015})}\BibitemShut
  {NoStop}%
\bibitem [{\citenamefont {Yoon}\ \emph {et~al.}(2018)\citenamefont {Yoon},
  \citenamefont {Dimitriadis}, \citenamefont {Soundararajan}, \citenamefont
  {Caspers}, \citenamefont {Genoud}, \citenamefont {Alberti}, \citenamefont
  {de~Rijk},\ and\ \citenamefont {Ansermet}}]{Yoon2018}%
  \BibitemOpen
  \bibfield  {author} {\bibinfo {author} {\bibfnamefont {D.}~\bibnamefont
  {Yoon}}, \bibinfo {author} {\bibfnamefont {A.~I.}\ \bibnamefont
  {Dimitriadis}}, \bibinfo {author} {\bibfnamefont {M.}~\bibnamefont
  {Soundararajan}}, \bibinfo {author} {\bibfnamefont {C.}~\bibnamefont
  {Caspers}}, \bibinfo {author} {\bibfnamefont {J.}~\bibnamefont {Genoud}},
  \bibinfo {author} {\bibfnamefont {S.}~\bibnamefont {Alberti}}, \bibinfo
  {author} {\bibfnamefont {E.}~\bibnamefont {de~Rijk}},\ and\ \bibinfo {author}
  {\bibfnamefont {J.-P.}\ \bibnamefont {Ansermet}},\ }\bibfield  {title}
  {\bibinfo {title} {High-field liquid-state dynamic nuclear polarization in
  microliter samples},\ }\href@noop {} {\bibfield  {journal} {\bibinfo
  {journal} {Anal. Chem.}\ }\textbf {\bibinfo {volume} {90}},\ \bibinfo {pages}
  {5620–5626} (\bibinfo {year} {2018})}\BibitemShut {NoStop}%
\bibitem [{\citenamefont {Corzilius}(2020)}]{Corzilius2020}%
  \BibitemOpen
  \bibfield  {author} {\bibinfo {author} {\bibfnamefont {B.}~\bibnamefont
  {Corzilius}},\ }\bibfield  {title} {\bibinfo {title} {High-field dynamic
  nuclear polarization},\ }\href@noop {} {\bibfield  {journal} {\bibinfo
  {journal} {Annual Rev. Phys. Chem.}\ }\textbf {\bibinfo {volume} {71}},\
  \bibinfo {pages} {143–170} (\bibinfo {year} {2020})}\BibitemShut {NoStop}%
\bibitem [{\citenamefont {Eills}\ \emph {et~al.}(2023)\citenamefont {Eills},
  \citenamefont {Budker}, \citenamefont {Cavagnero}, \citenamefont {Chekmenev},
  \citenamefont {Elliott}, \citenamefont {Jannin}, \citenamefont {Lesage},
  \citenamefont {Matysik}, \citenamefont {Meersmann}, \citenamefont {Prisner},
  \citenamefont {Reimer}, \citenamefont {Yang},\ and\ \citenamefont
  {Koptyug}}]{Eills2023}%
  \BibitemOpen
  \bibfield  {author} {\bibinfo {author} {\bibfnamefont {J.}~\bibnamefont
  {Eills}}, \bibinfo {author} {\bibfnamefont {D.}~\bibnamefont {Budker}},
  \bibinfo {author} {\bibfnamefont {S.}~\bibnamefont {Cavagnero}}, \bibinfo
  {author} {\bibfnamefont {E.~Y.}\ \bibnamefont {Chekmenev}}, \bibinfo {author}
  {\bibfnamefont {S.~J.}\ \bibnamefont {Elliott}}, \bibinfo {author}
  {\bibfnamefont {S.}~\bibnamefont {Jannin}}, \bibinfo {author} {\bibfnamefont
  {A.}~\bibnamefont {Lesage}}, \bibinfo {author} {\bibfnamefont
  {J.}~\bibnamefont {Matysik}}, \bibinfo {author} {\bibfnamefont
  {T.}~\bibnamefont {Meersmann}}, \bibinfo {author} {\bibfnamefont
  {T.}~\bibnamefont {Prisner}}, \bibinfo {author} {\bibfnamefont {J.~A.}\
  \bibnamefont {Reimer}}, \bibinfo {author} {\bibfnamefont {H.}~\bibnamefont
  {Yang}},\ and\ \bibinfo {author} {\bibfnamefont {I.~V.}\ \bibnamefont
  {Koptyug}},\ }\bibfield  {title} {\bibinfo {title} {Spin hyperpolarization in
  modern magnetic resonance},\ }\href@noop {} {\bibfield  {journal} {\bibinfo
  {journal} {C. R. Phys.}\ }\textbf {\bibinfo {volume} {123}},\ \bibinfo
  {pages} {1417} (\bibinfo {year} {2023})}\BibitemShut {NoStop}%
\bibitem [{\citenamefont {Fischer}\ \emph {et~al.}(2013)\citenamefont
  {Fischer}, \citenamefont {Bretschneider}, \citenamefont {London},
  \citenamefont {Budker}, \citenamefont {Gershoni},\ and\ \citenamefont
  {Frydman}}]{Fischer2013}%
  \BibitemOpen
  \bibfield  {author} {\bibinfo {author} {\bibfnamefont {R.}~\bibnamefont
  {Fischer}}, \bibinfo {author} {\bibfnamefont {C.~O.}\ \bibnamefont
  {Bretschneider}}, \bibinfo {author} {\bibfnamefont {P.}~\bibnamefont
  {London}}, \bibinfo {author} {\bibfnamefont {D.}~\bibnamefont {Budker}},
  \bibinfo {author} {\bibfnamefont {D.}~\bibnamefont {Gershoni}},\ and\
  \bibinfo {author} {\bibfnamefont {L.}~\bibnamefont {Frydman}},\ }\bibfield
  {title} {\bibinfo {title} {Bulk nuclear polarization enhanced at room
  temperature by optical pumping (eslac)},\ }\href@noop {} {\bibfield
  {journal} {\bibinfo  {journal} {Phys. Rev. Lett.}\ }\textbf {\bibinfo
  {volume} {111}},\ \bibinfo {pages} {057601} (\bibinfo {year}
  {2013})}\BibitemShut {NoStop}%
\bibitem [{\citenamefont {Wang}\ \emph {et~al.}(2013)\citenamefont {Wang},
  \citenamefont {Shin}, \citenamefont {Avalos}, \citenamefont {Seltzer},
  \citenamefont {Budker}, \citenamefont {Pines},\ and\ \citenamefont
  {Bajaj}}]{Wang2013}%
  \BibitemOpen
  \bibfield  {author} {\bibinfo {author} {\bibfnamefont {H.-J.}\ \bibnamefont
  {Wang}}, \bibinfo {author} {\bibfnamefont {C.~S.}\ \bibnamefont {Shin}},
  \bibinfo {author} {\bibfnamefont {C.~E.}\ \bibnamefont {Avalos}}, \bibinfo
  {author} {\bibfnamefont {S.~J.}\ \bibnamefont {Seltzer}}, \bibinfo {author}
  {\bibfnamefont {D.}~\bibnamefont {Budker}}, \bibinfo {author} {\bibfnamefont
  {A.}~\bibnamefont {Pines}},\ and\ \bibinfo {author} {\bibfnamefont {V.~S.}\
  \bibnamefont {Bajaj}},\ }\bibfield  {title} {\bibinfo {title} {Sensitive
  magnetic control of ensemble nuclear spin hyperpolarization in diamond},\
  }\href@noop {} {\bibfield  {journal} {\bibinfo  {journal} {Nat. Comm.}\
  }\textbf {\bibinfo {volume} {4}},\ \bibinfo {pages} {1940} (\bibinfo {year}
  {2013})}\BibitemShut {NoStop}%
\bibitem [{\citenamefont {Alvarez}\ \emph {et~al.}(2015)\citenamefont
  {Alvarez}, \citenamefont {Bretschneider}, \citenamefont {Fisher},
  \citenamefont {London}, \citenamefont {Kanda}, \citenamefont {Onoda},
  \citenamefont {Isoya}, \citenamefont {Gershoni},\ and\ \citenamefont
  {Frydman}}]{Alvarez2015}%
  \BibitemOpen
  \bibfield  {author} {\bibinfo {author} {\bibfnamefont {G.~A.}\ \bibnamefont
  {Alvarez}}, \bibinfo {author} {\bibfnamefont {C.~O.}\ \bibnamefont
  {Bretschneider}}, \bibinfo {author} {\bibfnamefont {R.}~\bibnamefont
  {Fisher}}, \bibinfo {author} {\bibfnamefont {P.}~\bibnamefont {London}},
  \bibinfo {author} {\bibfnamefont {H.}~\bibnamefont {Kanda}}, \bibinfo
  {author} {\bibfnamefont {S.}~\bibnamefont {Onoda}}, \bibinfo {author}
  {\bibfnamefont {J.}~\bibnamefont {Isoya}}, \bibinfo {author} {\bibfnamefont
  {D.}~\bibnamefont {Gershoni}},\ and\ \bibinfo {author} {\bibfnamefont
  {L.}~\bibnamefont {Frydman}},\ }\bibfield  {title} {\bibinfo {title} {Local
  and bulk $^{13}$c hyperpolarization in nitrogen-vacancy-centred diamonds at
  variable fields and orientations},\ }\href@noop {} {\bibfield  {journal}
  {\bibinfo  {journal} {Nat. Comm.}\ }\textbf {\bibinfo {volume} {6}},\
  \bibinfo {pages} {8456} (\bibinfo {year} {2015})}\BibitemShut {NoStop}%
\bibitem [{\citenamefont {King}\ \emph {et~al.}(2015)\citenamefont {King},
  \citenamefont {Jeong}, \citenamefont {Vassiliou}, \citenamefont {Shin},
  \citenamefont {Page}, \citenamefont {Avalos}, \citenamefont {Wang},\ and\
  \citenamefont {Pines}}]{King2015}%
  \BibitemOpen
  \bibfield  {author} {\bibinfo {author} {\bibfnamefont {J.~P.}\ \bibnamefont
  {King}}, \bibinfo {author} {\bibfnamefont {K.}~\bibnamefont {Jeong}},
  \bibinfo {author} {\bibfnamefont {C.~C.}\ \bibnamefont {Vassiliou}}, \bibinfo
  {author} {\bibfnamefont {C.~S.}\ \bibnamefont {Shin}}, \bibinfo {author}
  {\bibfnamefont {R.~H.}\ \bibnamefont {Page}}, \bibinfo {author}
  {\bibfnamefont {C.~E.}\ \bibnamefont {Avalos}}, \bibinfo {author}
  {\bibfnamefont {H.-J.}\ \bibnamefont {Wang}},\ and\ \bibinfo {author}
  {\bibfnamefont {A.}~\bibnamefont {Pines}},\ }\bibfield  {title} {\bibinfo
  {title} {Room-temperature in situ nuclear spin hyperpolarization from
  optically pumped nitrogen vacancy centres in diamond},\ }\href@noop {}
  {\bibfield  {journal} {\bibinfo  {journal} {Nat. Comm.}\ }\textbf {\bibinfo
  {volume} {6}},\ \bibinfo {pages} {8965} (\bibinfo {year} {2015})}\BibitemShut
  {NoStop}%
\bibitem [{\citenamefont {Scheuer}\ \emph {et~al.}(2016)\citenamefont
  {Scheuer}, \citenamefont {Schwartz}, \citenamefont {Chen}, \citenamefont
  {Schulze-Sünninghausen}, \citenamefont {Carl}, \citenamefont {Hofer},
  \citenamefont {Retzker}, \citenamefont {Sumiya}, \citenamefont {Isoya},
  \citenamefont {Luy}, \citenamefont {Plenio}, \citenamefont {Naydenov},\ and\
  \citenamefont {Jelezko}}]{Scheuer2016}%
  \BibitemOpen
  \bibfield  {author} {\bibinfo {author} {\bibfnamefont {J.}~\bibnamefont
  {Scheuer}}, \bibinfo {author} {\bibfnamefont {I.}~\bibnamefont {Schwartz}},
  \bibinfo {author} {\bibfnamefont {Q.}~\bibnamefont {Chen}}, \bibinfo {author}
  {\bibfnamefont {D.}~\bibnamefont {Schulze-Sünninghausen}}, \bibinfo {author}
  {\bibfnamefont {P.}~\bibnamefont {Carl}}, \bibinfo {author} {\bibfnamefont
  {P.}~\bibnamefont {Hofer}}, \bibinfo {author} {\bibfnamefont
  {A.}~\bibnamefont {Retzker}}, \bibinfo {author} {\bibfnamefont
  {H.}~\bibnamefont {Sumiya}}, \bibinfo {author} {\bibfnamefont
  {J.}~\bibnamefont {Isoya}}, \bibinfo {author} {\bibfnamefont
  {B.}~\bibnamefont {Luy}}, \bibinfo {author} {\bibfnamefont {M.~B.}\
  \bibnamefont {Plenio}}, \bibinfo {author} {\bibfnamefont {B.}~\bibnamefont
  {Naydenov}},\ and\ \bibinfo {author} {\bibfnamefont {F.}~\bibnamefont
  {Jelezko}},\ }\bibfield  {title} {\bibinfo {title} {Optically induced dynamic
  nuclear spin polarisation in diamond},\ }\href@noop {} {\bibfield  {journal}
  {\bibinfo  {journal} {New J. Phys.}\ }\textbf {\bibinfo {volume} {18}},\
  \bibinfo {pages} {013040} (\bibinfo {year} {2016})}\BibitemShut {NoStop}%
\bibitem [{\citenamefont {Ajoy}\ \emph {et~al.}(2018)\citenamefont {Ajoy},
  \citenamefont {Liu}, \citenamefont {Nazaryan}, \citenamefont {Lv},
  \citenamefont {Zangara}, \citenamefont {Safvati}, \citenamefont {Wang},
  \citenamefont {Arnold}, \citenamefont {Li}, \citenamefont {Lin},
  \citenamefont {Raghavan}, \citenamefont {Druga}, \citenamefont {Dhomkar},
  \citenamefont {Pagliero}, \citenamefont {Reimer}, \citenamefont {Suter},
  \citenamefont {Meriles},\ and\ \citenamefont {Pines}}]{Ajoy2018}%
  \BibitemOpen
  \bibfield  {author} {\bibinfo {author} {\bibfnamefont {A.}~\bibnamefont
  {Ajoy}}, \bibinfo {author} {\bibfnamefont {K.}~\bibnamefont {Liu}}, \bibinfo
  {author} {\bibfnamefont {R.}~\bibnamefont {Nazaryan}}, \bibinfo {author}
  {\bibfnamefont {X.}~\bibnamefont {Lv}}, \bibinfo {author} {\bibfnamefont
  {P.~R.}\ \bibnamefont {Zangara}}, \bibinfo {author} {\bibfnamefont
  {B.}~\bibnamefont {Safvati}}, \bibinfo {author} {\bibfnamefont
  {G.}~\bibnamefont {Wang}}, \bibinfo {author} {\bibfnamefont {D.}~\bibnamefont
  {Arnold}}, \bibinfo {author} {\bibfnamefont {G.}~\bibnamefont {Li}}, \bibinfo
  {author} {\bibfnamefont {A.}~\bibnamefont {Lin}}, \bibinfo {author}
  {\bibfnamefont {P.}~\bibnamefont {Raghavan}}, \bibinfo {author}
  {\bibfnamefont {E.}~\bibnamefont {Druga}}, \bibinfo {author} {\bibfnamefont
  {S.}~\bibnamefont {Dhomkar}}, \bibinfo {author} {\bibfnamefont
  {D.}~\bibnamefont {Pagliero}}, \bibinfo {author} {\bibfnamefont {J.~A.}\
  \bibnamefont {Reimer}}, \bibinfo {author} {\bibfnamefont {D.}~\bibnamefont
  {Suter}}, \bibinfo {author} {\bibfnamefont {C.~A.}\ \bibnamefont {Meriles}},\
  and\ \bibinfo {author} {\bibfnamefont {A.}~\bibnamefont {Pines}},\ }\bibfield
   {title} {\bibinfo {title} {Orientation-independent room temperature optical
  $^{13}$ hyperpolarization in powdered diamond},\ }\href@noop {} {\bibfield
  {journal} {\bibinfo  {journal} {Sci. Adv.}\ }\textbf {\bibinfo {volume}
  {4}},\ \bibinfo {pages} {eaar5492} (\bibinfo {year} {2018})}\BibitemShut
  {NoStop}%
\bibitem [{\citenamefont {Parker}\ \emph {et~al.}(2019)\citenamefont {Parker},
  \citenamefont {Jeong}, \citenamefont {Avalos}, \citenamefont {Hausmann},
  \citenamefont {Vassiliou}, \citenamefont {Pines},\ and\ \citenamefont
  {King}}]{Parker2019}%
  \BibitemOpen
  \bibfield  {author} {\bibinfo {author} {\bibfnamefont {A.~J.}\ \bibnamefont
  {Parker}}, \bibinfo {author} {\bibfnamefont {K.}~\bibnamefont {Jeong}},
  \bibinfo {author} {\bibfnamefont {C.~E.}\ \bibnamefont {Avalos}}, \bibinfo
  {author} {\bibfnamefont {B.~J.~M.}\ \bibnamefont {Hausmann}}, \bibinfo
  {author} {\bibfnamefont {C.~C.}\ \bibnamefont {Vassiliou}}, \bibinfo {author}
  {\bibfnamefont {A.}~\bibnamefont {Pines}},\ and\ \bibinfo {author}
  {\bibfnamefont {J.~P.}\ \bibnamefont {King}},\ }\bibfield  {title} {\bibinfo
  {title} {Optically pumped dynamic nuclear hyperpolarization in
  $^{13}$-enriched diamond},\ }\href@noop {} {\bibfield  {journal} {\bibinfo
  {journal} {Phys. Rev. B}\ }\textbf {\bibinfo {volume} {100}},\ \bibinfo
  {pages} {041203(R)} (\bibinfo {year} {2019})}\BibitemShut {NoStop}%
\bibitem [{\citenamefont {Scheuer}\ and\ \citenamefont
  {Naydenov}(2020)}]{Scheuer2020}%
  \BibitemOpen
  \bibfield  {author} {\bibinfo {author} {\bibfnamefont {J.}~\bibnamefont
  {Scheuer}}\ and\ \bibinfo {author} {\bibfnamefont {B.}~\bibnamefont
  {Naydenov}},\ }\bibfield  {title} {\bibinfo {title} {Dynamic nuclear
  polarization in diamond},\ }\href@noop {} {\bibfield  {journal} {\bibinfo
  {journal} {Semiconductors and Semimetals}\ }\textbf {\bibinfo {volume}
  {103}},\ \bibinfo {pages} {277–293} (\bibinfo {year} {2020})}\BibitemShut
  {NoStop}%
\bibitem [{\citenamefont {Ajoy}\ \emph {et~al.}(2020)\citenamefont {Ajoy},
  \citenamefont {Nazaryan}, \citenamefont {Druga}, \citenamefont {Liu},
  \citenamefont {Aguilar}, \citenamefont {Han}, \citenamefont {Gierth},
  \citenamefont {Oon}, \citenamefont {Safvati}, \citenamefont {Tsang},
  \citenamefont {Walton}, \citenamefont {Suter}, \citenamefont {Meriles},
  \citenamefont {Reimer},\ and\ \citenamefont {Pines}}]{Ajoy2020}%
  \BibitemOpen
  \bibfield  {author} {\bibinfo {author} {\bibfnamefont {A.}~\bibnamefont
  {Ajoy}}, \bibinfo {author} {\bibfnamefont {R.}~\bibnamefont {Nazaryan}},
  \bibinfo {author} {\bibfnamefont {E.}~\bibnamefont {Druga}}, \bibinfo
  {author} {\bibfnamefont {K.}~\bibnamefont {Liu}}, \bibinfo {author}
  {\bibfnamefont {A.}~\bibnamefont {Aguilar}}, \bibinfo {author} {\bibfnamefont
  {B.}~\bibnamefont {Han}}, \bibinfo {author} {\bibfnamefont {M.}~\bibnamefont
  {Gierth}}, \bibinfo {author} {\bibfnamefont {J.~T.}\ \bibnamefont {Oon}},
  \bibinfo {author} {\bibfnamefont {B.}~\bibnamefont {Safvati}}, \bibinfo
  {author} {\bibfnamefont {R.}~\bibnamefont {Tsang}}, \bibinfo {author}
  {\bibfnamefont {J.~H.}\ \bibnamefont {Walton}}, \bibinfo {author}
  {\bibfnamefont {D.}~\bibnamefont {Suter}}, \bibinfo {author} {\bibfnamefont
  {C.~A.}\ \bibnamefont {Meriles}}, \bibinfo {author} {\bibfnamefont {J.~A.}\
  \bibnamefont {Reimer}},\ and\ \bibinfo {author} {\bibfnamefont
  {A.}~\bibnamefont {Pines}},\ }\bibfield  {title} {\bibinfo {title} {Room
  temperature ``optical nanodiamond hyperpolarizer'': Physics, design, and
  operation},\ }\href@noop {} {\bibfield  {journal} {\bibinfo  {journal} {Rev.
  Sci. Instrum.}\ }\textbf {\bibinfo {volume} {91}},\ \bibinfo {pages} {023106}
  (\bibinfo {year} {2020})}\BibitemShut {NoStop}%
\bibitem [{\citenamefont {Sarkar}\ \emph {et~al.}(2022)\citenamefont {Sarkar},
  \citenamefont {Blankenship}, \citenamefont {Druga}, \citenamefont {Pillai},
  \citenamefont {Nirodi}, \citenamefont {Singh}, \citenamefont {Oddo},
  \citenamefont {Reshetikhin},\ and\ \citenamefont {Ajoy}}]{Sarkar2022}%
  \BibitemOpen
  \bibfield  {author} {\bibinfo {author} {\bibfnamefont {A.}~\bibnamefont
  {Sarkar}}, \bibinfo {author} {\bibfnamefont {B.}~\bibnamefont {Blankenship}},
  \bibinfo {author} {\bibfnamefont {E.}~\bibnamefont {Druga}}, \bibinfo
  {author} {\bibfnamefont {A.}~\bibnamefont {Pillai}}, \bibinfo {author}
  {\bibfnamefont {R.}~\bibnamefont {Nirodi}}, \bibinfo {author} {\bibfnamefont
  {S.}~\bibnamefont {Singh}}, \bibinfo {author} {\bibfnamefont
  {A.}~\bibnamefont {Oddo}}, \bibinfo {author} {\bibfnamefont {P.}~\bibnamefont
  {Reshetikhin}},\ and\ \bibinfo {author} {\bibfnamefont {A.}~\bibnamefont
  {Ajoy}},\ }\bibfield  {title} {\bibinfo {title} {Rapidly enhanced
  spin-polarization injection in an optically pumped spin ratchet},\ }\href
  {https://doi.org/10.1103/PhysRevApplied.18.034079} {\bibfield  {journal}
  {\bibinfo  {journal} {Phys. Rev. Appl.}\ }\textbf {\bibinfo {volume} {18}},\
  \bibinfo {pages} {034079} (\bibinfo {year} {2022})}\BibitemShut {NoStop}%
\bibitem [{\citenamefont {Gierth}\ \emph {et~al.}(2020)\citenamefont {Gierth},
  \citenamefont {Krespach}, \citenamefont {Shames}, \citenamefont {Raghavan},
  \citenamefont {Druga}, \citenamefont {Nunn}, \citenamefont {Torelli},
  \citenamefont {Nirodi}, \citenamefont {Le}, \citenamefont {Zhao},
  \citenamefont {Aguilar}, \citenamefont {Lv}, \citenamefont {Shen},
  \citenamefont {Meriles}, \citenamefont {Reimer}, \citenamefont {Zaitsev},
  \citenamefont {Pines}, \citenamefont {Shenderova},\ and\ \citenamefont
  {Ajoy}}]{Gierth2020}%
  \BibitemOpen
  \bibfield  {author} {\bibinfo {author} {\bibfnamefont {M.}~\bibnamefont
  {Gierth}}, \bibinfo {author} {\bibfnamefont {V.}~\bibnamefont {Krespach}},
  \bibinfo {author} {\bibfnamefont {A.~I.}\ \bibnamefont {Shames}}, \bibinfo
  {author} {\bibfnamefont {P.}~\bibnamefont {Raghavan}}, \bibinfo {author}
  {\bibfnamefont {E.}~\bibnamefont {Druga}}, \bibinfo {author} {\bibfnamefont
  {N.}~\bibnamefont {Nunn}}, \bibinfo {author} {\bibfnamefont {M.}~\bibnamefont
  {Torelli}}, \bibinfo {author} {\bibfnamefont {R.}~\bibnamefont {Nirodi}},
  \bibinfo {author} {\bibfnamefont {S.}~\bibnamefont {Le}}, \bibinfo {author}
  {\bibfnamefont {R.}~\bibnamefont {Zhao}}, \bibinfo {author} {\bibfnamefont
  {A.}~\bibnamefont {Aguilar}}, \bibinfo {author} {\bibfnamefont
  {X.}~\bibnamefont {Lv}}, \bibinfo {author} {\bibfnamefont {M.}~\bibnamefont
  {Shen}}, \bibinfo {author} {\bibfnamefont {C.~A.}\ \bibnamefont {Meriles}},
  \bibinfo {author} {\bibfnamefont {J.~A.}\ \bibnamefont {Reimer}}, \bibinfo
  {author} {\bibfnamefont {A.}~\bibnamefont {Zaitsev}}, \bibinfo {author}
  {\bibfnamefont {A.}~\bibnamefont {Pines}}, \bibinfo {author} {\bibfnamefont
  {O.}~\bibnamefont {Shenderova}},\ and\ \bibinfo {author} {\bibfnamefont
  {A.}~\bibnamefont {Ajoy}},\ }\bibfield  {title} {\bibinfo {title} {Enhanced
  optical $^{13}$C hyperpolarization in diamond treated by high-temperature rapid
  thermal annealing},\ }\href@noop {} {\bibfield  {journal} {\bibinfo
  {journal} {Advanced Quantum Technologies}\ }\textbf {\bibinfo {volume} {3}},\
  \bibinfo {pages} {2000050} (\bibinfo {year} {2020})}\BibitemShut {NoStop}%
\bibitem [{\citenamefont {Bauch}\ \emph {et~al.}(2020)\citenamefont {Bauch},
	\citenamefont {Singh}, \citenamefont {Lee}, \citenamefont {Hart},
	\citenamefont {Schloss}, \citenamefont {Turner}, \citenamefont {Barry},
	\citenamefont {Pham}, \citenamefont {Bar-Gill}, \citenamefont {Yelin},\ and\
	\citenamefont {Walsworth}}]{Bauch2020}%
\BibitemOpen
\bibfield  {author} {\bibinfo {author} {\bibfnamefont {E.}~\bibnamefont {Bauch}},
	\bibinfo {author} {\bibfnamefont {S.}~\bibnamefont {Singh}}, \bibinfo {author}
	{\bibfnamefont {J.}~\bibnamefont {Lee}}, \bibinfo {author} {\bibfnamefont {C.~A.}\ \bibnamefont {Hart}},
	\bibinfo {author} {\bibfnamefont {J.~M.}\ \bibnamefont {Schloss}}, \bibinfo
	{author} {\bibfnamefont {M.~J.}\ \bibnamefont {Turner}}, \bibinfo {author}
	{\bibfnamefont {J.~F.}\ \bibnamefont {Barry}}, \bibinfo {author}
	{\bibfnamefont {L.~M.}\ \bibnamefont {Pham}}, \bibinfo {author}
	{\bibfnamefont {N.}~\bibnamefont {Bar-Gill}}, \bibinfo {author}
	{\bibfnamefont {S.~F.}\ \bibnamefont {Yelin}},\ and\ \bibinfo {author}
	{\bibfnamefont {R.~L.}\ \bibnamefont {Walsworth}},\ }\bibinfo {title}{Decoherence of ensembles of nitrogen-vacancy centers in diamond. }\href
{https://doi.org/10.1103/PhysRevB.102.134210} {\bibfield  {journal}
	{\bibinfo  {journal} {Phys. Rev. B}\ }\textbf {\bibinfo {volume} {102}},\
	\bibinfo {pages} {134210} (\bibinfo {year} {2020})}\BibitemShut {NoStop}
\bibitem [{\citenamefont {Nalbach}\ and\ \citenamefont
  {Thorwart}(2010)}]{Nalbach2010}%
  \BibitemOpen
  \bibfield  {author} {\bibinfo {author} {\bibfnamefont {P.}~\bibnamefont
  {Nalbach}}\ and\ \bibinfo {author} {\bibfnamefont {M.}~\bibnamefont
  {Thorwart}},\ }\bibfield  {title} {\bibinfo {title} {Competition between
  relaxation and external driving in the dissipative landau–zener problem},\
  }\href {https://doi.org/https://doi.org/10.1016/j.chemphys.2010.05.007}
  {\bibfield  {journal} {\bibinfo  {journal} {Chemical Physics}\ }\textbf
  {\bibinfo {volume} {375}},\ \bibinfo {pages} {234} (\bibinfo {year}
  {2010})},\ \bibinfo {note} {stochastic processes in Physics and Chemistry (in
  honor of Peter Hänggi)}\BibitemShut {NoStop}%
\bibitem [{\citenamefont {Kavtanyuk}\ \emph {et~al.}(2022)\citenamefont
  {Kavtanyuk}, \citenamefont {Lee}, \citenamefont {Oh}, \citenamefont {Jeong},\
  and\ \citenamefont {Shim}}]{molecules27051700}%
  \BibitemOpen
  \bibfield  {author} {\bibinfo {author} {\bibfnamefont {V.~V.}\ \bibnamefont
  {Kavtanyuk}}, \bibinfo {author} {\bibfnamefont {H.~J.}\ \bibnamefont {Lee}},
  \bibinfo {author} {\bibfnamefont {S.}~\bibnamefont {Oh}}, \bibinfo {author}
  {\bibfnamefont {K.}~\bibnamefont {Jeong}},\ and\ \bibinfo {author}
  {\bibfnamefont {J.~H.}\ \bibnamefont {Shim}},\ }\bibfield  {title} {\bibinfo
  {title} {Optical dynamic nuclear polarization of $^{13}$c spins in diamond at
  a low field with multi-tone microwave irradiation},\ }\href
  {https://doi.org/10.3390/molecules27051700} {\bibfield  {journal} {\bibinfo
  {journal} {Molecules}\ }\textbf {\bibinfo {volume} {27}},\ \bibinfo {pages}
  {1700} (\bibinfo {year} {2022})}\BibitemShut {NoStop}%
\bibitem [{\citenamefont {Abobeih}\ \emph {et~al.}(2019)\citenamefont
  {Abobeih}, \citenamefont {Randall}, \citenamefont {Bradley}, \citenamefont
  {Bartling}, \citenamefont {Bakker}, \citenamefont {Degen}, \citenamefont
  {Markham}, \citenamefont {Twitchen},\ and\ \citenamefont
  {Taminiau}}]{Abobeih2019}%
  \BibitemOpen
  \bibfield  {author} {\bibinfo {author} {\bibfnamefont {M.~H.}\ \bibnamefont
  {Abobeih}}, \bibinfo {author} {\bibfnamefont {J.}~\bibnamefont {Randall}},
  \bibinfo {author} {\bibfnamefont {C.~E.}\ \bibnamefont {Bradley}}, \bibinfo
  {author} {\bibfnamefont {H.~P.}\ \bibnamefont {Bartling}}, \bibinfo {author}
  {\bibfnamefont {M.~A.}\ \bibnamefont {Bakker}}, \bibinfo {author}
  {\bibfnamefont {M.~J.}\ \bibnamefont {Degen}}, \bibinfo {author}
  {\bibfnamefont {M.}~\bibnamefont {Markham}}, \bibinfo {author} {\bibfnamefont
  {D.~J.}\ \bibnamefont {Twitchen}},\ and\ \bibinfo {author} {\bibfnamefont
  {T.~H.}\ \bibnamefont {Taminiau}},\ }\bibfield  {title} {\bibinfo {title}
  {Atomic-scale imaging of a 27-nuclear-spin cluster using a quantum sensor},\
  }\href {https://doi.org/10.1038/s41586-019-1834-7} {\bibfield  {journal}
  {\bibinfo  {journal} {Nature}\ }\textbf {\bibinfo {volume} {576}},\ \bibinfo
  {pages} {411} (\bibinfo {year} {2019})}\BibitemShut {NoStop}%
\bibitem [{\citenamefont {Wenckebach}(2016)}]{Wenckebach2016}%
  \BibitemOpen
  \bibfield  {author} {\bibinfo {author} {\bibfnamefont {T.}~\bibnamefont
  {Wenckebach}},\ }\href@noop {} {\emph {\bibinfo {title} {Essentials of
  Dynamic Nuclear Polarization}}}\ (\bibinfo  {publisher} {Spindrift
  Publications, The Netherlands},\ \bibinfo {year} {2016})\BibitemShut
  {NoStop}%
\bibitem [{\citenamefont {Can}\ \emph {et~al.}(2018)\citenamefont {Can},
  \citenamefont {McKay}, \citenamefont {Weber}, \citenamefont {Yang},
  \citenamefont {Dubroca}, \citenamefont {van Tol}, \citenamefont {Hill},\ and\
  \citenamefont {Griffin}}]{Can2018}%
  \BibitemOpen
  \bibfield  {author} {\bibinfo {author} {\bibfnamefont {T.~V.}\ \bibnamefont
  {Can}}, \bibinfo {author} {\bibfnamefont {J.~E.}\ \bibnamefont {McKay}},
  \bibinfo {author} {\bibfnamefont {R.~T.}\ \bibnamefont {Weber}}, \bibinfo
  {author} {\bibfnamefont {C.}~\bibnamefont {Yang}}, \bibinfo {author}
  {\bibfnamefont {T.}~\bibnamefont {Dubroca}}, \bibinfo {author} {\bibfnamefont
  {J.}~\bibnamefont {van Tol}}, \bibinfo {author} {\bibfnamefont
  {S.}~\bibnamefont {Hill}},\ and\ \bibinfo {author} {\bibfnamefont {R.~G.}\
  \bibnamefont {Griffin}},\ }\bibfield  {title} {\bibinfo {title}
  {Frequency-swept integrated and stretched solid effect dynamic nuclear
  polarization},\ }\href {https://doi.org/10.1021/acs.jpclett.8b01002}
  {\bibfield  {journal} {\bibinfo  {journal} {J. Phys. Chem. Lett.}\ }\textbf
  {\bibinfo {volume} {9}},\ \bibinfo {pages} {3187} (\bibinfo {year}
  {2018})}\BibitemShut {NoStop}%
\bibitem [{\citenamefont {Tetienne}\ \emph {et~al.}(2012)\citenamefont
  {Tetienne}, \citenamefont {Rondin}, \citenamefont {Spinicelli}, \citenamefont
  {Chipaux}, \citenamefont {Debuisschert}, \citenamefont {Roch},\ and\
  \citenamefont {Jacques}}]{Tetienne2012}%
  \BibitemOpen
  \bibfield  {author} {\bibinfo {author} {\bibfnamefont {J.-P.}\ \bibnamefont
  {Tetienne}}, \bibinfo {author} {\bibfnamefont {L.}~\bibnamefont {Rondin}},
  \bibinfo {author} {\bibfnamefont {P.}~\bibnamefont {Spinicelli}}, \bibinfo
  {author} {\bibfnamefont {M.}~\bibnamefont {Chipaux}}, \bibinfo {author}
  {\bibfnamefont {T.}~\bibnamefont {Debuisschert}}, \bibinfo {author}
  {\bibfnamefont {J.-F.}\ \bibnamefont {Roch}},\ and\ \bibinfo {author}
  {\bibfnamefont {V.}~\bibnamefont {Jacques}},\ }\bibfield  {title} {\bibinfo
  {title} {Magnetic-field-dependent photodynamics of single nv defects in
  diamond: an application to qualitative all-optical magnetic imaging},\ }\href
  {https://doi.org/10.1088/1367-2630/14/10/103033} {\bibfield  {journal}
  {\bibinfo  {journal} {New Journal of Physics}\ }\textbf {\bibinfo {volume}
  {14}},\ \bibinfo {pages} {103033} (\bibinfo {year} {2012})}\BibitemShut
  {NoStop}%
\bibitem [{\citenamefont {Acosta}\ \emph {et~al.}(2010)\citenamefont {Acosta},
  \citenamefont {Bauch}, \citenamefont {Ledbetter}, \citenamefont {Waxman},
  \citenamefont {Bouchard},\ and\ \citenamefont {Budker}}]{Acosta2010}%
  \BibitemOpen
  \bibfield  {author} {\bibinfo {author} {\bibfnamefont {V.~M.}\ \bibnamefont
  {Acosta}}, \bibinfo {author} {\bibfnamefont {E.}~\bibnamefont {Bauch}},
  \bibinfo {author} {\bibfnamefont {M.~P.}\ \bibnamefont {Ledbetter}}, \bibinfo
  {author} {\bibfnamefont {A.}~\bibnamefont {Waxman}}, \bibinfo {author}
  {\bibfnamefont {L.-S.}\ \bibnamefont {Bouchard}},\ and\ \bibinfo {author}
  {\bibfnamefont {D.}~\bibnamefont {Budker}},\ }\href
  {https://doi.org/10.1103/PhysRevLett.104.070801} {\bibfield  {journal}
  {\bibinfo  {journal} {Phys. Rev. Lett.}\ }\textbf {\bibinfo {volume} {104}},\
  \bibinfo {pages} {070801} (\bibinfo {year} {2010})}\BibitemShut {NoStop}%
\bibitem [{\citenamefont {Shim}\ \emph {et~al.}(2013)\citenamefont {Shim},
  \citenamefont {Niemeyer}, \citenamefont {Zhang},\ and\ \citenamefont
  {Suter}}]{Shim2013}%
  \BibitemOpen
  \bibfield  {author} {\bibinfo {author} {\bibfnamefont {J.~H.}\ \bibnamefont
  {Shim}}, \bibinfo {author} {\bibfnamefont {I.}~\bibnamefont {Niemeyer}},
  \bibinfo {author} {\bibfnamefont {J.}~\bibnamefont {Zhang}},\ and\ \bibinfo
  {author} {\bibfnamefont {D.}~\bibnamefont {Suter}},\ }\bibfield  {title}
  {\bibinfo {title} {Room-temperature high-speed nuclear-spin quantum memory in
  diamond},\ }\href {https://doi.org/10.1103/PhysRevA.87.012301} {\bibfield
  {journal} {\bibinfo  {journal} {Phys. Rev. A}\ }\textbf {\bibinfo {volume}
  {87}},\ \bibinfo {pages} {012301} (\bibinfo {year} {2013})}\BibitemShut
  {NoStop}%
\end{thebibliography}
\providecommand{\noopsort}[1]{}\providecommand{\singleletter}[1]{#1}%

\end{document}